\documentclass[]{spie}  

\newcommand{\ukrts}{$\mu$K$\sqrt{\textrm{s}}$}

\usepackage{amsmath,amsfonts,amssymb}
\usepackage{graphicx}
\usepackage[colorlinks=true, allcolors=blue]{hyperref}
\usepackage{siunitx}
\usepackage{booktabs}
\usepackage{caption} 
\usepackage{color}
\usepackage{tablefootnote}
\usepackage[normalem]{ulem}

\title{Observing low elevation sky and the CMB Cold Spot with BICEP3 at the South Pole}

\author[a,b]{Jae Hwan Kang}
\affil[a]{Department of Physics, California Institute of Technology, Pasadena, California 91125, USA}
\affil[b]{Department of Physics, Stanford University, Stanford, California 94305, USA}
\author[n]{P.~A.~R.~Ade}
\author[c,d]{Z.~Ahmed}
\affil[c]{SLAC National Accelerator Laboratory, 2575 Sand Hill Road, Menlo Park, CA 94025}
\affil[d]{Kavli Institute for Particle Astrophysics and Cosmology, Stanford University, 452 Lomita Mall, Stanford, CA 94305}
\author[e]{M.~Amiri}
\affil[e]{Department of Physics and Astronomy, University of British Columbia, Vancouver, British Columbia, V6T 1Z1, Canada}
\author[f,g]{D.~Barkats}
\affil[f]{Institut Laue-Langevin, 38042 Grenoble Cedex 9, France}
\affil[g]{Center for Astrophysics $\vert$ Harvard \& Smithsonian, Cambridge, MA 02138, U.S.A}
\author[a]{R.~Basu Thakur}
\author[h]{C.~A.~Bischoff}
\affil[h]{Department of Physics, University of Cincinnati, Cincinnati, Ohio 45221, USA}
\author[a,i]{J.~J.~Bock}
\affil[i]{Jet Propulsion Laboratory, Pasadena, California 91109, USA}
\author[g]{H.~Boenish}
\author[j]{E.~Bullock}
\affil[j]{Minnesota Institute for Astrophysics, University of Minnesota, Minneapolis, 55455, USA}
\author[k]{V.~Buza}
\affil[k]{Kavli Institute for Cosmological Physics, University of Chicago, Chicago, IL 60637, USA}
\author[l]{J.~R.~Cheshire}
\affil[l]{School of Physics and Astronomy, University of Minnesota, Minneapolis, 55455, USA}
\author[m]{J.~Connors}
\affil[m]{National Institute of Standards and Technology, Boulder, Colorado 80305, USA}
\author[g]{J.~Cornelison}
\author[l]{M.~Crumrine}
\author[d,c,b]{A.~Cukierman}
\affil[n]{School of Physics and Astronomy, Cardiff University, Cardiff, CF24 3AA, United Kingdom}
\author[m]{E.~Denison}
\author[g]{M.~Dierickx}
\author[o]{L.~Duband}
\affil[o]{Service des Basses Temp\'{e}ratures, Commissariat \`{a} lEnergie Atomique, 38054 Grenoble, France}
\author[g]{M.~Eiben}
\author[e]{S.~Fatigoni}
\author[p,q]{J.~P.~Filippini}
\affil[p]{Department of Physics, University of Illinois at Urbana-Champaign, Urbana, Illinois 61801}
\affil[q]{Department of Astronomy, University of Illinois at Urbana-Champaign, Urbana, Illinois 61801, USA}
\author[l]{S.~Fliescher}
\author[d,b]{N.~Goeckner-Wald}
\author[g]{D.~C.~Goldfinger}
\author[b]{J.~A.~Grayson}
\author[g]{P.~Grimes}
\author[l]{G.~Hall}
\author[e]{M.~Halpern}
\author[g]{S.~A.~Harrison}
\author[c,d]{S.~Henderson}
\author[a,i]{S.~R.~Hildebrandt}
\author[m]{G.~C.~Hilton}
\author[m]{J.~Hubmayr}
\author[a]{H.~Hui}
\author[c,d,b,m]{K.~D.~Irwin}
\author[k]{K.~S.~Karkare}
\author[b]{E.~Karpel}
\author[a]{S.~Kefeli}
\author[b]{S.~A.~Kernasovskiy}
\author[g,r]{J.~M.~Kovac}
\affil[r]{Department of Physics, Harvard University, Cambridge, MA 02138, USA}
\author[b,c,d]{C.~L.~Kuo}
\author[l]{K.~Lau}
\author[k]{E.~M.~Leitch}
\author[i]{K.~G.~Megerian}
\author[a]{L.~Minutolo}
\author[a]{L.~Moncelsi}
\author[l]{Y.~Nakato}
\author[s]{T.~Namikawa}
\affil[s]{Department of Applied Mathematics and Theoretical Physics, University of Cambridge, Cambridge CB3 0WA, UK}
\author[a,i]{H.~T.~Nguyen}
\author[a,i]{R.~O'Brient}
\author[b]{R.~W.~Ogburn~IV}
\author[h]{S.~Palladino}
\author[l]{N.~Precup}
\author[o]{T.~Prouve}
\author[k,l]{C.~Pryke}
\author[g]{B.~Racine}
\author[m]{C.~D.~Reintsema}
\author[g]{S.~Richter}
\author[a]{A.~Schillaci}
\author[g]{B.~L.~Schmitt}
\author[l]{R.~Schwarz}
\author[j]{C.~D.~Sheehy}
\author[a]{A.~Soliman}
\author[g,r]{T.~St.~Germaine}
\author[a]{B.~Steinbach}
\author[n]{R.~V.~Sudiwala}
\author[t]{G.~P.~Teply}
\affil[t]{Department of Physics, University of California at San Diego, La Jolla, California 92093, USA}
\author[d,b]{K.~L.~Thompson}
\author[b]{J.~E.~Tolan}
\author[n]{C.~Tucker}
\author[i]{A.~D.~Turner}
\author[p,q]{C.~Umilt\`{a}}
\author[k]{A.~G.~Vieregg}
\author[a]{A.~Wandui}
\author[i]{A.~C.~Weber}
\author[e]{D.~V.~Wiebe}
\author[l]{J.~Willmert}
\author[g,r]{C.~L.~Wong}
\author[c,d,b]{W.~L.~K.~Wu}
\author[b]{E.~Yang}
\author[b,c,d]{K.~W.~Yoon}
\author[d,c,b]{E.~Young}
\author[b]{C.~Yu}
\author[g]{L.~Zeng}
\author[a]{C.~Zhang}
\author[a]{S.~Zhang}

\authorinfo{Corresponding author: J. Kang, jkang7@caltech.edu}

\begin{document} 
	\maketitle
	
	\begin{abstract}
		 BICEP3 is a 520 mm aperture on-axis refracting telescope at the South Pole, which observes the polarization of the cosmic microwave background (CMB) at 95 GHz to search for the B-mode signal from inflationary gravitational waves. In addition to this main target, we have developed a low-elevation observation strategy to extend coverage of the Southern sky at the South Pole, where BICEP3 can quickly achieve degree-scale E-mode measurements over a large area. An interesting E-mode measurement is probing a potential polarization anomaly around the CMB Cold Spot. During the austral summer seasons of 2018-19 and 2019-20, BICEP3 observed the sky with a flat mirror to redirect the beams to various low elevation ranges. The preliminary data analysis shows degree-scale E-modes measured with high signal-to-noise ratio.
	\end{abstract}
	
	
	\section{INTRODUCTION}
	\label{sec:intro}  
	
	The cosmic microwave background (CMB) is weakly linearly polarized by Thomson scattering at the last scattering surface (LSS). The linear polarization can be decomposed to curl-free E-modes and divergence-free B-modes \cite{Bmode1997_PRD.55.7368,Bmode1997_PRD.55.Seljak_Zaldarriaga}. Stochastic density perturbations produce E-modes and the anisotropy of the E-mode fluctuations have been measured at various angular scales: from the first detection on a small sky patch by DASI at degree-angular scales\cite{DASI_Leitch_2002,Kovac2002} and from the full sky survey by Planck\cite{Planck_2018_I_Overview}. Classes of inflationary theories predict stochastic gravitational wave background from tensor perturbations of the spacetime metric and imprints B-modes as well as E-modes on the CMB. Detection of B-modes from the CMB is a strong validation of the inflation paradigm. The power spectrum of the inflationary B-mode CMB polarization is expected to peak around degree angular scales. The amplitude of the power spectrum is related to the energy scale of the inflation and characterized by the ratio of the amplitude of the tensor perturbation to that of the scalar perturbation, called tensor-to-scalar ratio $r$\cite{Kamionkowski2016}.
	
	The faint inflationary B-mode signal can be confused with Galactic foregrounds such as synchrotron radiation and polarized dust emission. Furthermore, the brighter CMB E-modes can be leaked into B-modes via gravitational lensing by mass distributions throughout the line of the sight to the LSS. The BICEP/Keck Array series of telescopes have been observing the CMB polarization with receivers at multiple frequencies to separate the CMB components from Galactic foregrounds which exhibit different spectral dependencies from the CMB. Continued efforts are being made in both detector count and frequency coverage with the development and deployment of the BICEP Array\cite{Moncelsi2020}.
	
	BICEP3 is a third-generation telescope on the BICEP mount in the Dark Sector Lab (DSL) at the South Pole. It was designed to obtain deep polarization map at 95 GHz on a small patch of the sky in the Southern hemisphere where Galactic dust emission is relatively weak\cite{Ahmed14}. BICEP3 was deployed in the austral summer of 2014-15 (hereafter, we refer all seasons as austral seasons). After a winter engineering season in 2015, BICEP3 has been accumulating science data since the 2016 winter season\cite{Grayson16}. In 2017, BICEP3 underwent improvements including the replacement of infrared filters and underperforming detector modules, allowing it to achieve an on-sky instantaneous sensitivity of 6.7 $\mathrm{\mu K_{cmb}\sqrt{s}}$\cite{Kang18SPIE}. It has kept the same observing configuration ever since. The inclusion of accumulated BICEP3 data at 95 GHz along with data from other receivers in the BICEP/Keck series at other frequencies on the same sky patch is expected to improve the constraints on $r$ by a factor of two compared to the previously published result of $\sigma(r)=0.020$ from the data accumulated to 2015 season\cite{BK-X_BK15,Moncelsi2020}.
	
	As BICEP3 continues to collect polarization data in our main observing patch, we also explored opportunities for BICEP3 to leverage its fast degree-angular scale polarization mapping speed. BICEP3 can measure degree-scale E-modes with high signal-to-noise ratio in a wide patch of sky available from the Southern hemisphere. An extended sky coverage can be useful to understand non-isotropic non-Gaussian polarized foregrounds and to study other large scale E-mode science cases.
	
	While the statistics of temperature and E-mode anisotropies from Planck data has enabled precision measurements of the standard $\Lambda$CDM cosmological parameters\cite{Planck_2018_VI_Cosmological_Parameters}, anomalies in the CMB which deviate from the random Gaussian fluctuations predicted by the standard cosmology are interesting subjects to study. A particularly interesting anomaly visible to BICEP3 with an extended coverage observation strategy is the CMB Cold Spot, which was first detected in the WMAP temperature map\cite{WMAP_1st_yr_ColdSpot}. The Cold Spot contributes to the anomalous statistics of Spherical Mexican Hat Wavelet (SMHW) coefficients of the temperature map at the angular scales of around 250 arcminutes. A conservative estimate of the probability of finding such a feature in Gaussian simulations after taking the effect of \textit{a posteriori} selection into account is 1.85\%\cite{WMAP3yr_Cruz_2007}, and the Cold Spot was again identified in the Planck temperature map\cite{Planck_2013_XXIII_Isotropy_and_Statistics}.
	
	A comprehensive review on the Cold Spot including its possible explanations is provided by Ref.~\citenum{Vielva_CS_review_2010}. Aside from random statistical fluke from Gaussian fluctuations, plausible explanations include a cooling of the CMB photons through the Sachs-Wolfe (SW) effect from evolving gravitational potentials sourced by a collapsing texture at high redshift\cite{Texture_Turok_Spergel_1990}, or a supervoid at low redshift\cite{CS_voids_Inoue_2006}. A template fitting to the effect of texture showed compatibility with a collapsing texture at $z\sim 6$\cite{CS_texture_science_2007}. A galaxy survey along the line of sight to the Cold Spot identifies voids at $z<0.5$, but the estimated integrated SW effect fell short of the observed temperature drop at the Cold Spot\cite{CS_against_supervoid_Mackenzie_2017}.

	A follow-up polarization observation around the Cold Spot can help distinguish the texture hypothesis from random Gaussian fluctuation of the CMB, because the cooling of CMB photons by evolving gravitational potentials would not associate with CMB polarization, and observed E-modes around the Cold Spot would be less likely to be consistent with the $\Lambda$CDM CMB TE correlation. Ref.~\citenum{ColdSpot_FC13} developed a method to use polarization observation around the Cold Spot to distinguish these hypotheses. They require polarization sensitivity of 0.3 $\mathrm{\mu K\ deg}$ within a $20^\circ$ radius around the Cold Spot to achieve the significance level of 1.5\% at the power of the test of 0.5, with the most discriminating power coming from a $7^\circ$ radius.
	
	Plank has provided exquisite precision cosmology from full sky statistics of both temperature and E-mode anisotropies. However, in a small patch of the sky, the signal-to-noise in polarization is small, e.g. polarization sensitivity of 1.96 $\mathrm{\mu K_{cmb}\ deg}$ at 100 GHz\cite{Planck_2018_I_Overview}. An examination of Planck polarization data around the Cold Spot was not conclusive to test polarization anomaly due to limited polarzation sensitivity\cite{Planck_2018_VII_Isotropy_and_Statistics}. In its regular observing patch, BICEP3 can quickly reach the map depth deeper than the Planck polarization map, thus motivating observations around the Cold Spot for an initial BICEP3 extended observing campaign.
	
	In this proceeding, we present the performance of BICEP3 at low elevation ranges to demonstrate the extent of potential sky coverage at the South Pole. These measurements were conducted during the summer seasons of 2018-19 and 2019-20, outside the regular observing season. Section \ref{sec:instrument} gives an instrument overview that is relevant to the configuration of the low elevation observation. Section \ref{sec:observation} presents the observing strategy around the location of the Cold Spot. Section \ref{sec:performance} presents the performance of BICEP3 in this observation mode. The preliminary polarization maps and angular power spectra are shown.
	
	\section{Instrument Overview}
	\label{sec:instrument}

\begin{figure}[t]
		\centering
		\includegraphics[width=0.7\textwidth]{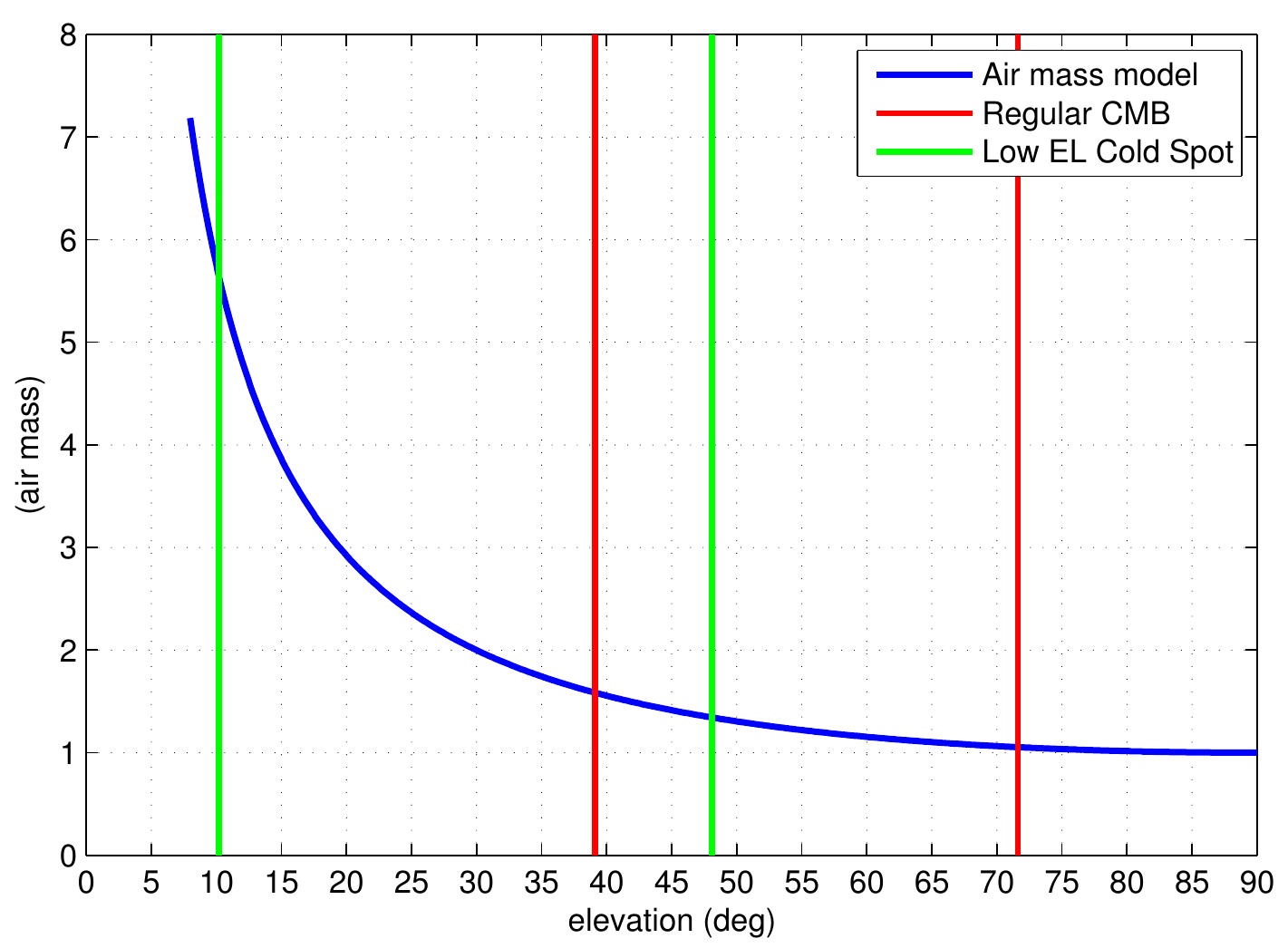}
		\caption[Airmass model as $\csc(el)$]{The air mass model as $\csc(\mathrm{EL})$ is shown in blue. The elevation ranges for the presented low elevation schedules and the regular CMB schedules are marked with green and red vertical lines, respectively.}
		\label{fig.cs_cmb_airmass}
	\end{figure}

	BICEP3 is a 520 mm aperture on-axis refracting telescope with its focal plane populated with about 2500 transition edge sensor (TES) bolometers coupled with polarization sensitive antenna arrays\cite{TES_BK}. A detailed BICEP3 instrument overview can be found in Ref.~\citenum{graysonthesis}. The antenna arrays collect incoming photon energy and transmit it to thermal islands through microstrips. The transmitted radiation energy deposits heat to the island and dissipates to the bolometers and then out of the island through thin legs to the thermal bath. To keep the TES bolometers at the negative electro-thermal feedback stage during observation, we determine the optimum TES bias points using typical observation loading conditions. The current detector design chose thermal conductance of the legs of the bolometer islands to have a safety factor of 2.5 for expected in-band on-sky loading per detector. The atmosphere is estimated to have about 10 times more power than the CMB at BICEP3's 95 GHz observing band across the elevation ranges of the regular CMB field at the South Pole\cite{HuiThesis}. The airmass along the line of sight becomes thicker as we point to lower elevation, closely modeled by a cosecant of the observing elevation. As Figure \ref{fig.cs_cmb_airmass} shows, the change in airmass is much steeper below $45^\circ$ elevation for this observing campaign compared to the elevation range for regular CMB observation. At the same boresight elevation, changing the boresight orientation can bring a detector at a low elevation to a high elevation. Instead of applying the same TES bias points for all boresight orientations, we tune the biases for each boresight orientation. Even with this retuning, detectors at the lowest elevations are still saturated over the safety margin chosen for regular CMB schedules due to atmospheric loading. We tested detector response at various elevation ranges and find the lower limit of observing elevation with the current detectors (see Figure \ref{fig.round2_overall_passfrac}).
	
	The bolometers are cooled to 250 mK by a three-stage helium sorption fridge\footnote{Chase Research Cryogenics Ltd., Sheeld, S10 5DL, UK (www.chasecryogenics.com)}. The focal plane unit (FPU) is housed on the sub-Kelvin structure inside the 4K stage. The optics tube is cryogenically cooled in vacuum to 50K and 4K stages by a PT415 Pulsetube cryocooler\footnote{Cryomech Inc., Syracuse, NY 13211, USA (www.cryomech.com)} through oxygen-free high-conductivity (OFHC) copper heat straps. The pulsetube coldhead is attached to the outermost vacuum shell. The cooling performance of the coldhead depends on the orientation with respect to gravity. Since the performance degrades quickly beyond $45^\circ$ away from the vertical, we do not tilt the receiver lower than $48^\circ$ elevation\cite{graysonthesis}. The upper limit of elevation is set by the range of the sector wheel friction drive. Therefore, the elevation axis of the mount is limited to $48^\circ < \mathrm{el} < 110^\circ$. The elevation axis of the mount is the elevation axis of the boresight. The corresponding elevation limits of the observing field is defined by the 28$^\circ$ field of view of BICEP3. At the low elevation limit, the observing field does not go below 34$^\circ$ elevation.
	
\begin{figure}[ht]
		\centering
		\includegraphics[width=0.66\textwidth]{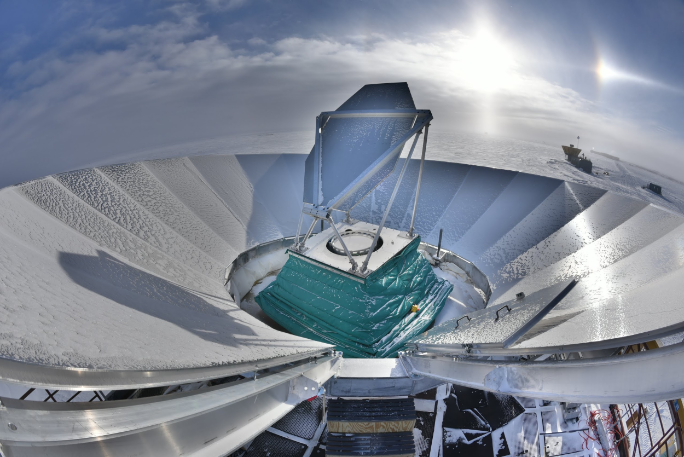}
		\caption[BICEP3 with the Far Field Flat on]{BICEP3 with the flat mirror on. The photo was taken when the ground shield was tipped to point the source mounted on the neighboring building for the far field beam calibration. For observation of on-sky sources, the ground shield is leveled.}
		\label{fig.B3_FFFlat_on}
	\end{figure}

	In order to reach low elevation ranges where the mount cannot directly point, we use a large flat aluminum mirror to redirect the beams. The mirror has been used for far field beam calibrations\cite{StGermaine2020} and absolute polarization response angle measurements\cite{Cornelison2020}. Designed to point sources mounted on the neighboring building, the BICEP3 mirror redirects all detector beams by nearly $90^\circ$. When the mount is pointed at zenith, the telescope boresight points near the horizon. With the attachment of the heavy mirror, we further limit the mount range of motion, keeping the boresight at less than $34^\circ$ in elevation. Due to the large field of view of BICEP3, the detectors cover enough elevation ranges for some of them to have overlap with regular CMB observing elevation. Figure \ref{fig.cs_cmb_airmass} includes the vertical lines which mark the elevation ranges for the presented low elevation observing schedules and the regular CMB schedules. Figure \ref{fig.B3_FFFlat_on} shows BICEP3 with the flat mirror on. We note that the forebaffle could not fit with the mirror simultaneously. A design with proper baffling will improve data quality of sky observation with fewer effects from spurious scattering.
	
	\section{OBSERVING STRATEGY}
	\label{sec:observation}

	The observing configuration requires station personnel to install and uninstall the flat mirror. To conserve our regular winter CMB observation, we conducted this observing campaign during the summer seasons of 2018-19 and 2019-20. Both campaigns fell within the 2019 calendar year, so we will refer them to early 2019 and late 2019 observations.
	
\begin{figure}[t]
		\centering
		\includegraphics[width=1\textwidth]{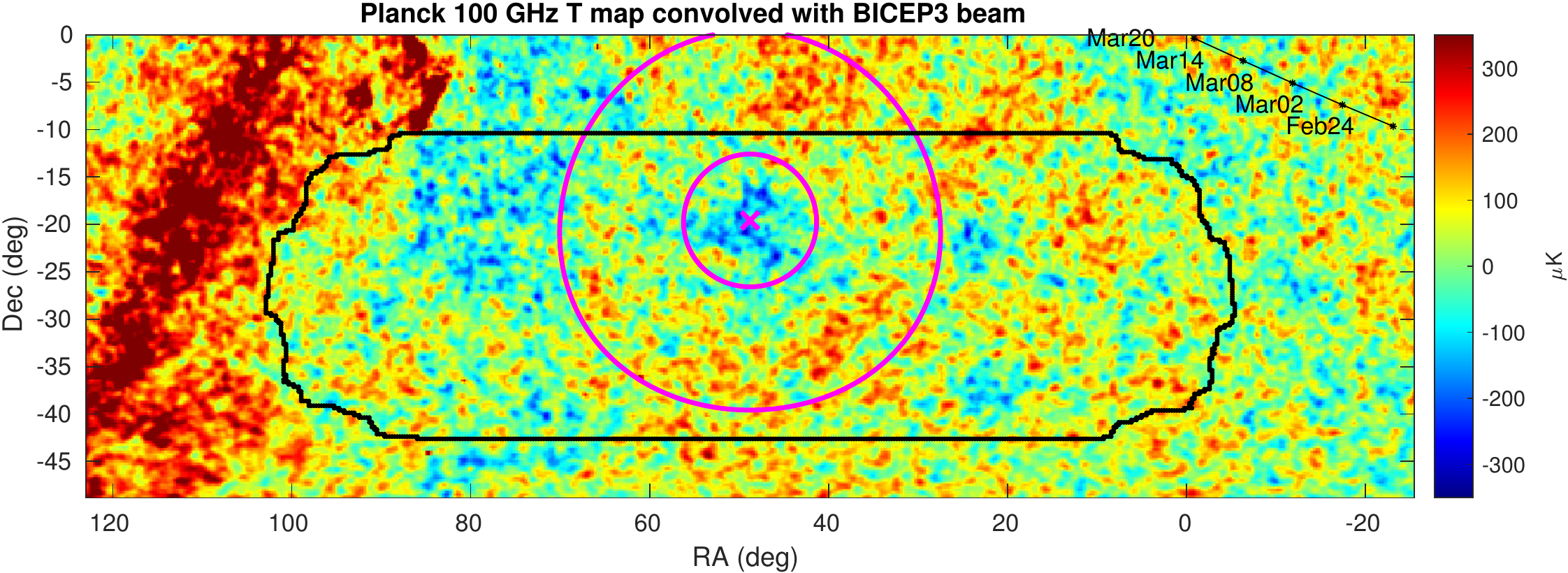}
		\caption[BICEP3 Cold Spot observation covarage plans]{Coverage plan plotted over the Planck 100 GHz temperature map convolved with the BICEP3 beam. The magenta cross indicates the location of the center of the Cold Spot, and the concentric circles indicate $7^\circ$ and $20^\circ$ radii regions around the Cold Spot. The circles look distorted due to the flat map projection. The path of the Sun is marked with a label every 6 days. The black boundary indicates the coverage by BICEP3 with a set of Cold Spot schedules.}
		\label{fig.plans_over_Planck_T}
	\end{figure}

	The CMB Cold Spot is centered at the Galactic coordinates $(l, b) = (210^\circ,-57^\circ)$ which in equatorial coordinates is $(RA,Dec) = (48.77^\circ,-19.58^\circ)$. At the South Pole, the Cold Spot sits around $20^\circ$ elevation. When we point boresight elevation too low, many detectors will be intercepted by the ground shield and other ground features. Detectors clearing the ground features may still get saturated due to thick atmosphere at low elevation. We have tested multiple elevation ranges. Figure \ref{fig.plans_over_Planck_T} shows the boundary of the coverage from a set of observing schedules. The contours are plotted over the Planck 100 GHz temperature map\footnote{https://pla.esac.esa.int/\#home} convolved with the BICEP3 beam. The magenta cross marks the center of the Cold Spot, with concentric circles of $7^\circ$ and $20^\circ$ radii. Since the observing patch is near the ecliptic, we need to consider the position of the Sun and the Moon. The path of the Sun is marked with a label every six days. The Sun approaches the patch during the late summer and can affect the data taken around that time. The Moon was far away during the observation. The bottom-right corner of the coverage has a small overlap with the regular CMB patch.
	
\begin{figure}[t]
	\centering
	\includegraphics[width=1\textwidth]{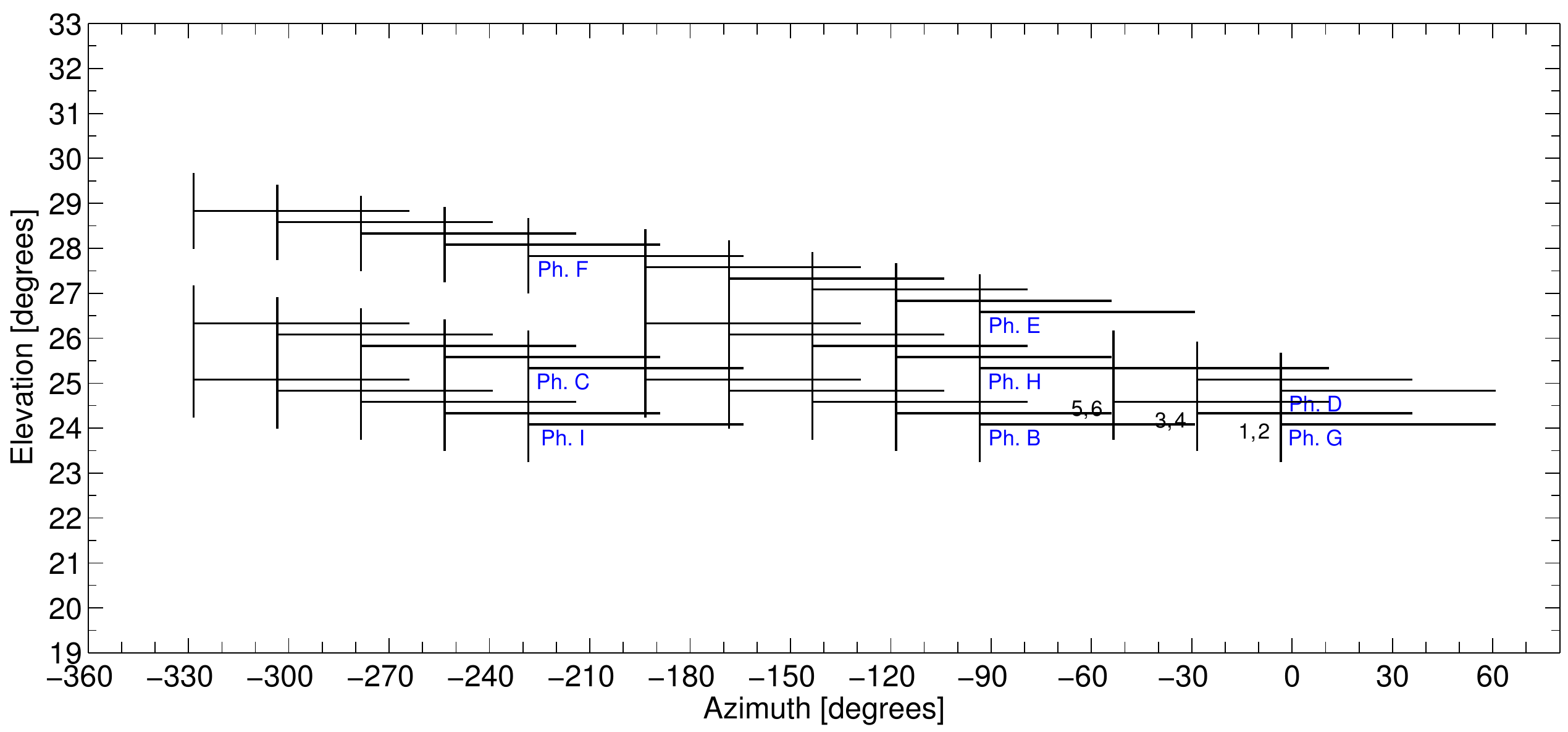}
	\caption[BICEP3 Cold Spot scan strategy - LM schedules]{Observing pattern of a three-day schedule at low elevation range in ground-based coordinates. Horizontal lines indicate the field scans and the vertical lines indicate the bracketing elevation nods. The telescope scans at a fixed elevation during each scanset. We observe two scansets before changing elevation.
	}
	\label{fig.scan_strategy_cs_v4}
\end{figure}

	Figure \ref{fig.scan_strategy_cs_v4} shows the observing pattern of a typical three-day schedule in ground-based coordinates. We adopt the same strategy as the regular CMB observing schedules\cite{Kang18SPIE}. The fundamental unit is a `scanset' which consists of 50 back-and-forth scans at a fixed elevation at a speed of $2.8^\circ/\mathrm{s}$ in a fixed azimuth range. At the start and end of each scanset are bracketing elevation nods to calibrate relative detector gains in airmass units. Each scanset takes about 50 minutes to complete, and the usable observing time excluding turnarounds and calibrations is around 33 minutes. Every other scansets, we move boresight elevation and adjust the observing azimuth center to follow drifting sky. Six or ten scansets are grouped as phases to which the deprojection technique for beam systematic mitigation is applied\cite{BK-III}. All phases observe the same patch instead of reserving one phase for the Galactic plane, as is done in regular CMB observing. We change boresight orientation for each three-day schedule to map polarization. The low observing schedules may be affected by ground objects. Future observing campaigns would adjust azimuth ranges to reduce the number of scansets affected by the South Pole Telescope (SPT) on the same building, which is presently visible around the azimuth range of -10 to -40 degrees at low elevation.
	
	Choosing the scansets that are less affected by the Sun and ground features, we obtain 336 good scansets from January-March 2019 schedules and 783 scansets from October-November 2019 schedules. The early 2019 schedules include scansets that are closer to the regular CMB field at higher elevation, while the late 2019 schedules are focused around the Cold Spot (See the boundary in Figure \ref{fig.plans_over_Planck_T}). This gives the usable integration time of about 8 days      from the early 2019 data, and 18 days from the late 2019 data.
	
	For regular CMB observation, we use optical star pointing to derive mount pointing parameters that may change in time. For the summer observation with the mirror on, we cannot run star pointing. For early 2019 data, we corrected boresight pointing by correlating the coadded temperature map with Planck map at various assumed pointing offsets. For late 2019 data, we performed Moon rasters to use the very well known location of the Moon to derive parameters of the mirror orientation.	
	
	\section{DETECTOR PERFORMANCE}
	\label{sec:performance}

\begin{figure}[tb]
	\centering
	\includegraphics[width=0.32\textwidth]{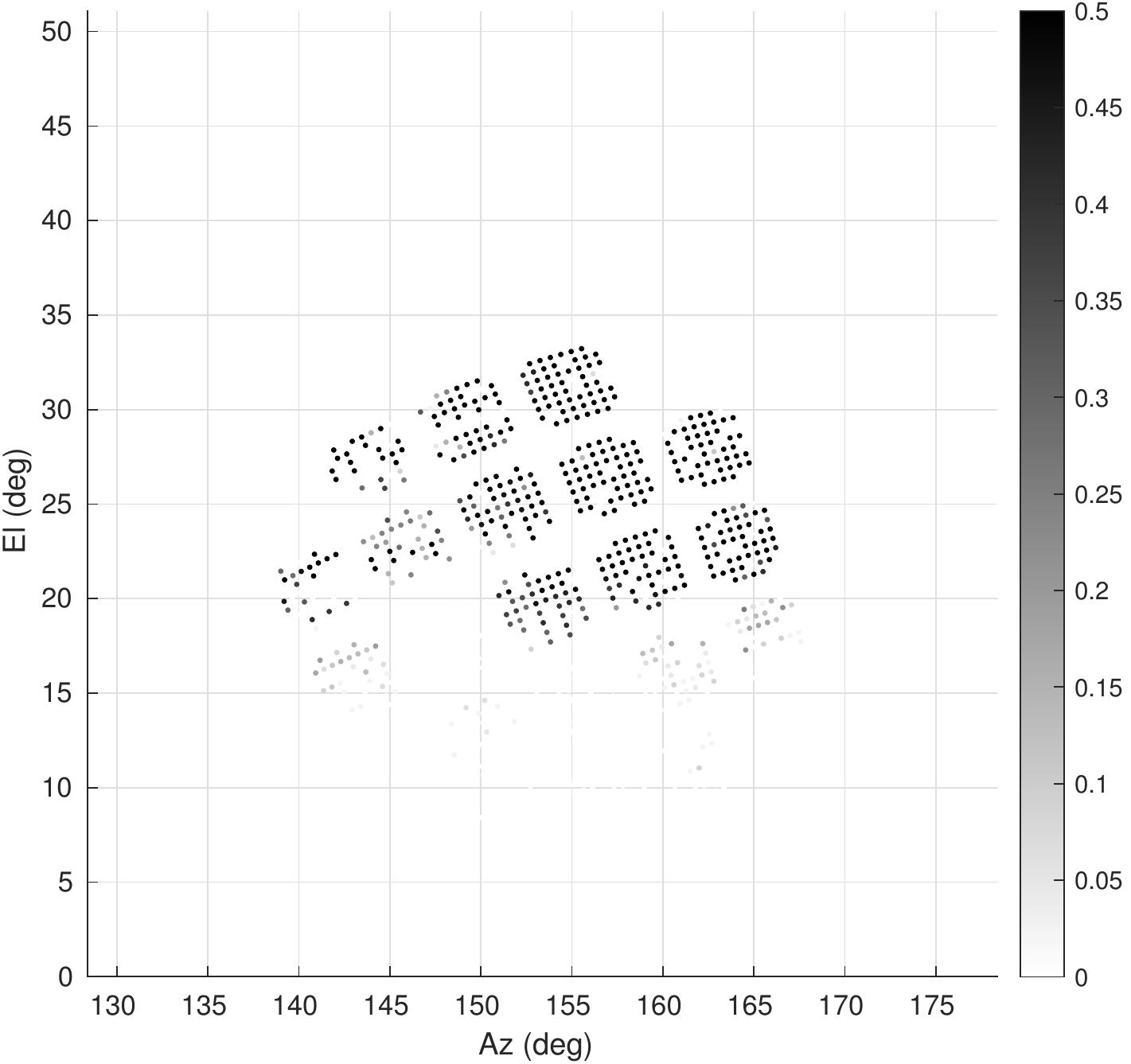}
	\includegraphics[width=0.32\textwidth]{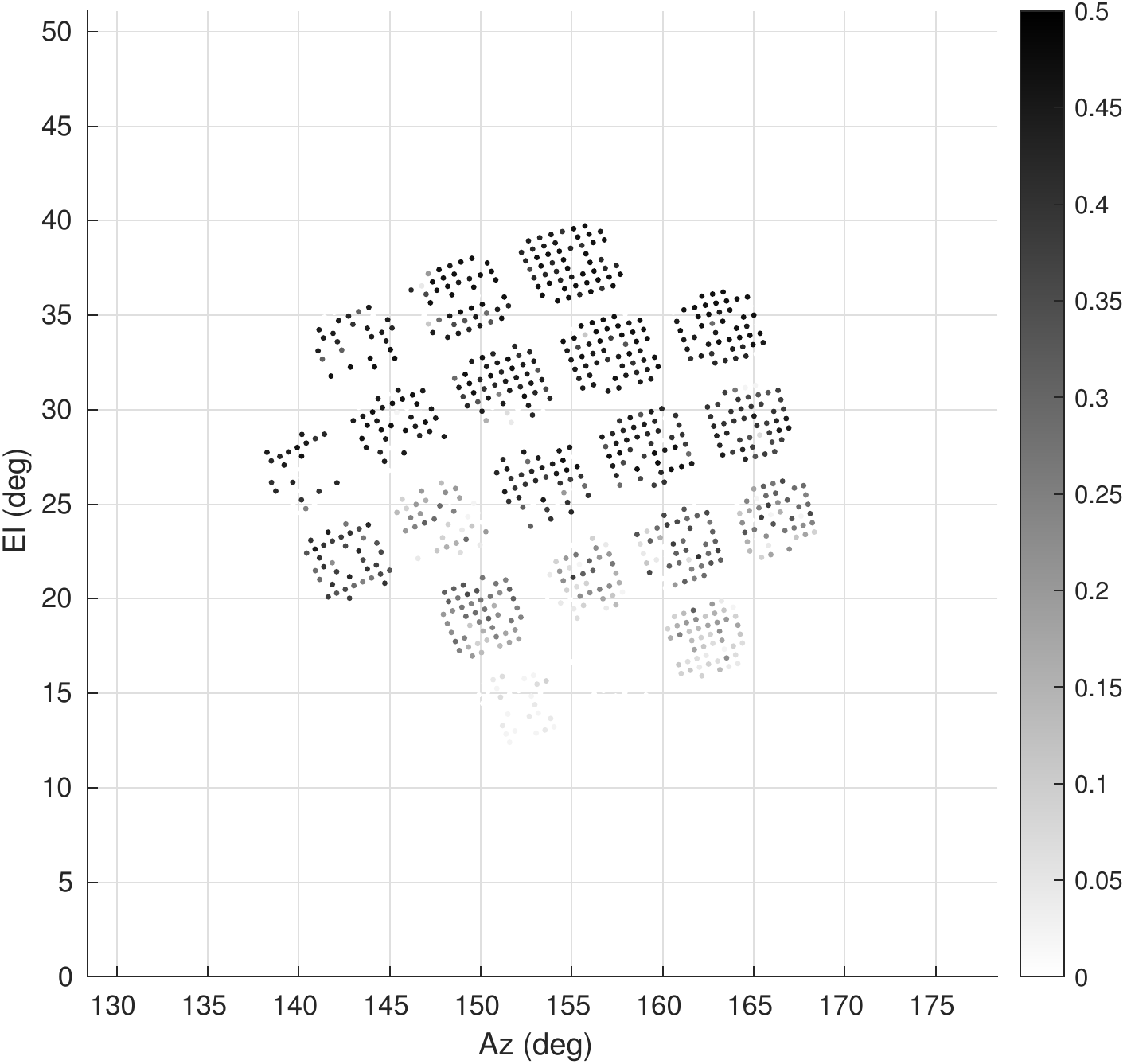}
	\includegraphics[width=0.32\textwidth]{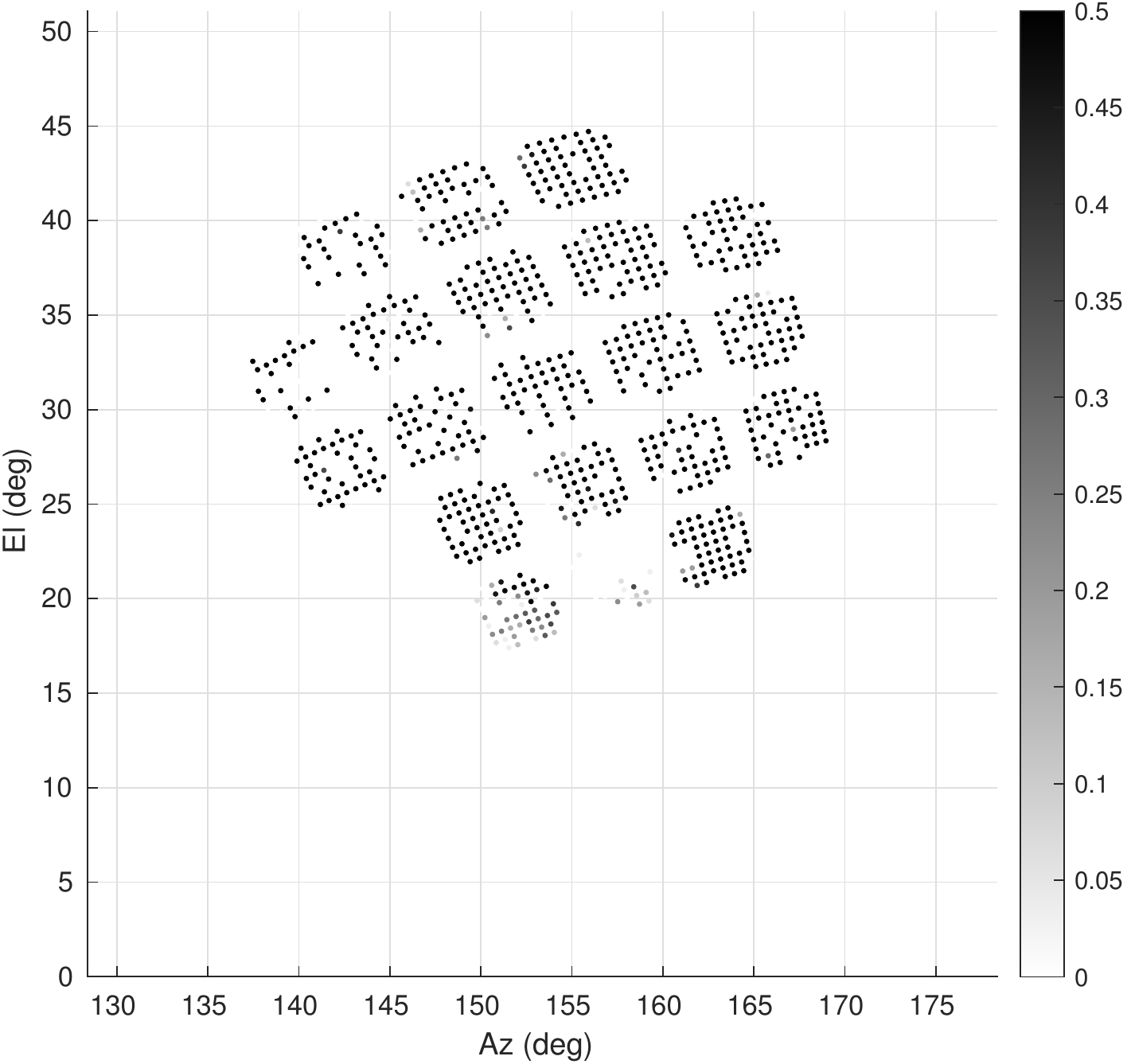}
	\caption[Round2 overall pass fractions for different observing schedules]{Pass fraction of detectors after data quality cuts\cite{Kang_thesis}. Schedules from different elevation centers with the same boresight orientation are shown. Detectors perform worse at lower elevation, and they get mostly saturated and stop working properly below EL of $15^\circ$.}
	\label{fig.round2_overall_passfrac}
\end{figure}

	\begin{figure} [t]
	\begin{center}
		\begin{tabular}{c}
			\includegraphics[width=.82\textwidth]{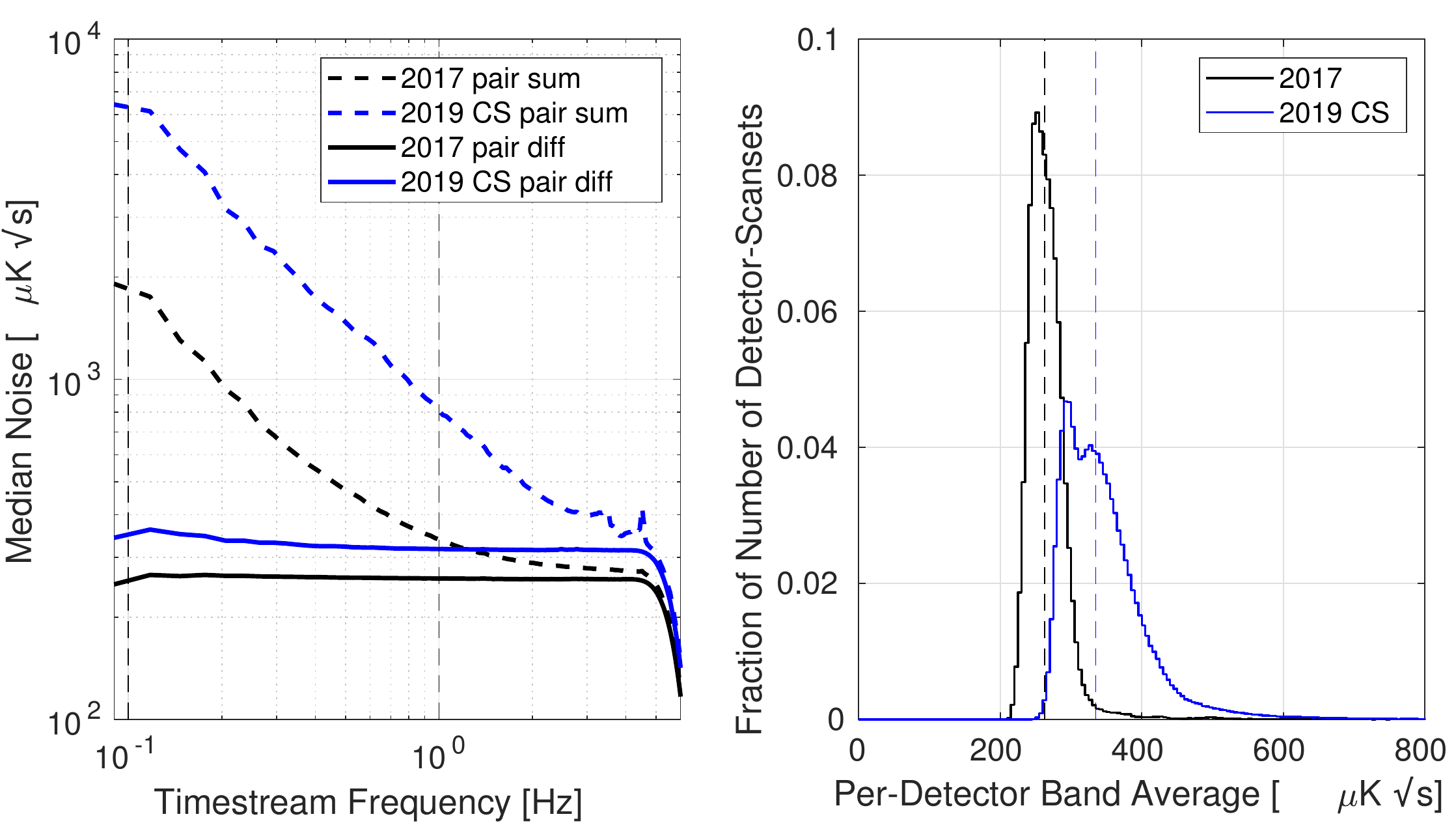}
		\end{tabular}
	\end{center}
	\caption[example] 
	{ \label{fig:NETspec} 
		\textit{Left}: Median per-detector noise spectra for BICEP3 2017 season data and 2019 Cold Spot observation data, from pair-summed and pair-differenced timestreams. The timestreams are subjected to a third-order polynomial filtering for each half scan to reject $1/f$ noise.
		\textit{Right}: Histogram of the per-detector per-scanset noise, applying 3rd-order polynomial timestream filtering and averaged across the 0.1-1 Hz science band. Median values are marked by vertical dashed lines, respectively. The 2019 Cold Spot data has a median 335 \ukrts, above the 265 \ukrts\, from 2017 data\cite{Kang18SPIE}.
	}
\end{figure}

\subsection{Detector Response}

Figure \ref{fig.round2_overall_passfrac} shows the pass fraction of detectors from the early 2019 schedules after data quality cuts\cite{Kang_thesis}. Schedules from different elevation centers with the same boresight orientation are shown. Detectors perform worse at lower elevation, and they get mostly saturated and stop working properly below elevation of $15^\circ$. The current design of the bolometers are not optimized for this high atmospheric loading condition (see Figure \ref{fig.cs_cmb_airmass}).

	\subsection{Timestream-based NET}	
	The detector performance can be assessed by estimating NET based on timestream data as we showed in previous proceedings\cite{Grayson16,Kang18SPIE}. For each scanset, we take pair-summed and pair-differenced timestream data and apply data quality cuts to keep only well-behaving detector pairs. We convert them to CMB temperature by calibrating the measured temperature map to the 100 GHz Planck temperature map. We take the power spectral density (PSD) of pair-differenced timestreams across each azimuthal scan and multiply by two to get per-detector estimate instead of per-pair estimate. The median spectra across all detectors and scansets are plotted on the left panel of Figure~\ref{fig:NETspec}. The timestreams are subjected to a third-order polynomial filtering for each half scan to reject $1/f$ noise. We find the mean across the usual science band 0.1-1 Hz per detector per scanset and make the histogram over each data set on the right panel of Figure~\ref{fig:NETspec}. The noise level is increased in the low elevation observation centered around $26^\circ$ elevation compared to a regular season with observing center around $55^\circ$ elevation. The 2019 Cold Spot data has a median 335 \ukrts, above the 265 \ukrts\, from 2017 data\cite{Kang18SPIE}.
	
	\subsection{Maps and Power Spectra}
	\begin{figure} [tb]
		\begin{center}
			\begin{tabular}{c} 
				\includegraphics[width=0.8\textwidth]{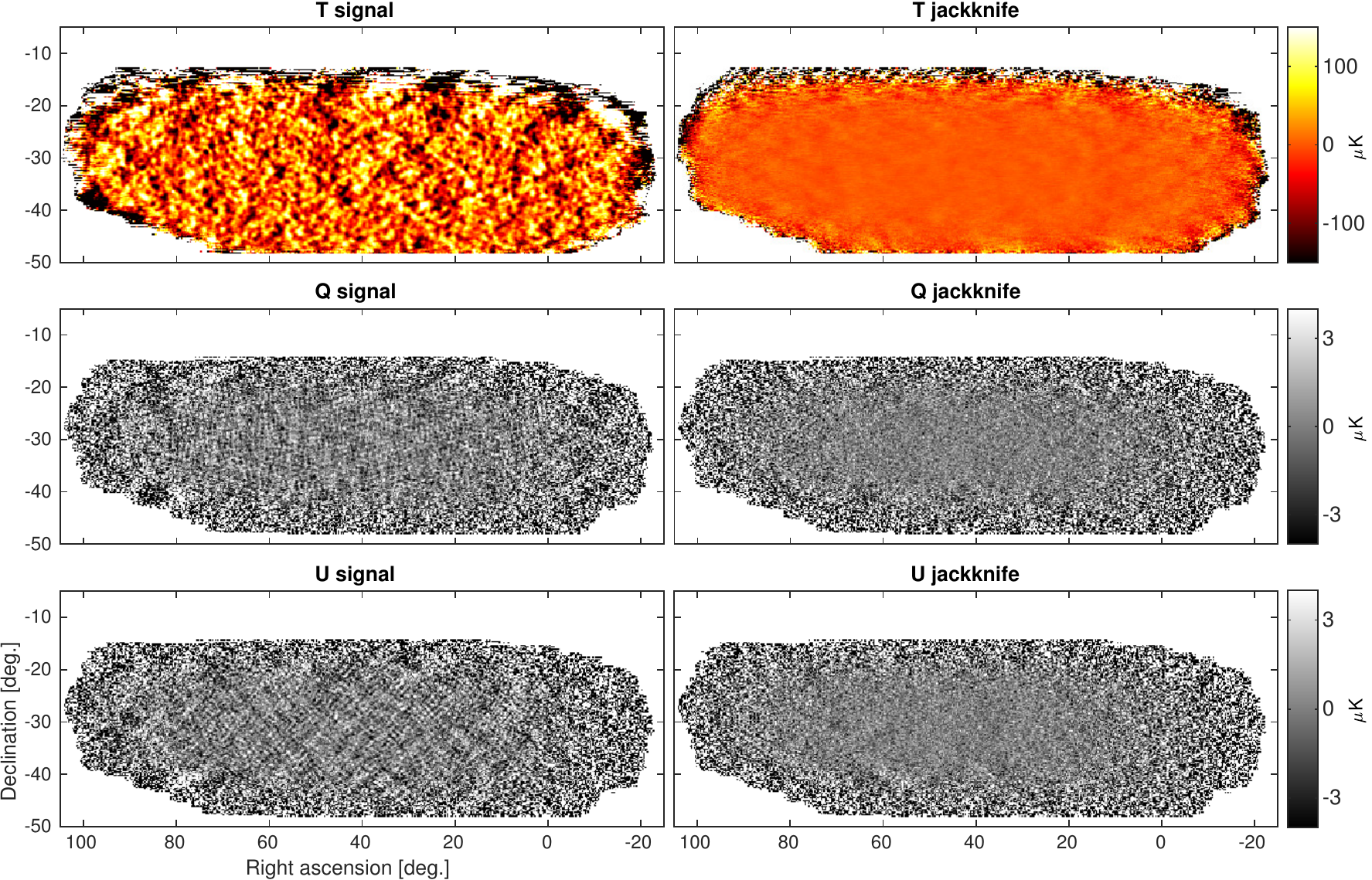}
			\end{tabular}
		\end{center}
		\caption[example] 
		{ \label{fig:B2019CS-I_II_tqumap} Temperature, Q and U polarization maps (\textit{Left}) and their noise estimates (\textit{Right}) from BICEP3 early and late 2019 season Cold Spot observations\cite{Kang_thesis}. The scan-direction jackknife maps are used as a proxy to show noise level.}
	\end{figure} 
	\begin{figure} [tb]
		\begin{center}
			\begin{tabular}{c}
				\includegraphics[width=0.8\textwidth]{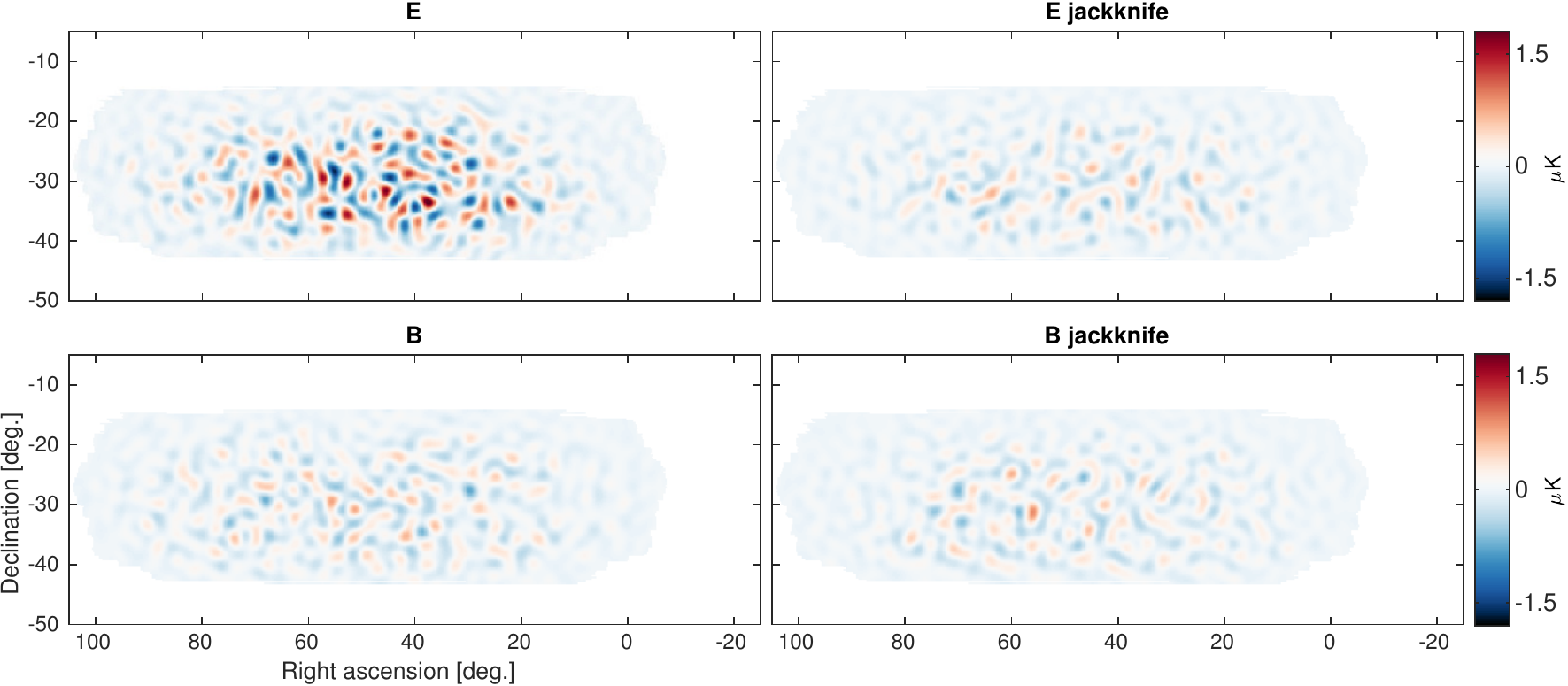}
			\end{tabular}
		\end{center}
		\caption[example] 
		{ \label{fig:B2019CS-II_ebmap} E-mode and B-mode maps representing the observed modes within $50<\ell<120$ from late 2019 season Cold Spot observations (\textit{Left}) and their noise estimates (\textit{Right})\cite{Kang_thesis}.
		Degree scale E-mode is detected at high signal-to-noise, while B-mode is noise-dominated.}
	\end{figure} 

	We apply the standard BICEP/Keck Array reduction pipeline\cite{BK-I} with modifications necessary to process the data taken with the flat mirror. Figure~\ref{fig:B2019CS-I_II_tqumap} shows the temperature, Q and U polarization maps and their noise estimates from the preliminary analysis of the early and late 2019 season Cold Spot observations combined\cite{Kang_thesis}. We show scan-direction jackknife maps as a proxy to noise. The Cold Spot feature is visible in the temperature map around the right ascension of $50^\circ$ and declination of $-20^\circ$. The characteristic $+$ and $\times$ patterns of E-modes from density fluctuations are visible in Q and U maps, respectively. Saturating detectors at low elevation through large air mass (Figure \ref{fig.round2_overall_passfrac}) results in poor map sensitivity at higher declination than the Cold Spot. Therefore, we need to deal with partial coverage of the area of interest around the Cold Spot for the test of polarization anomaly.

	\begin{figure} [tb]
	\begin{center}
		\begin{tabular}{c}
			\includegraphics[width=0.4\textwidth]{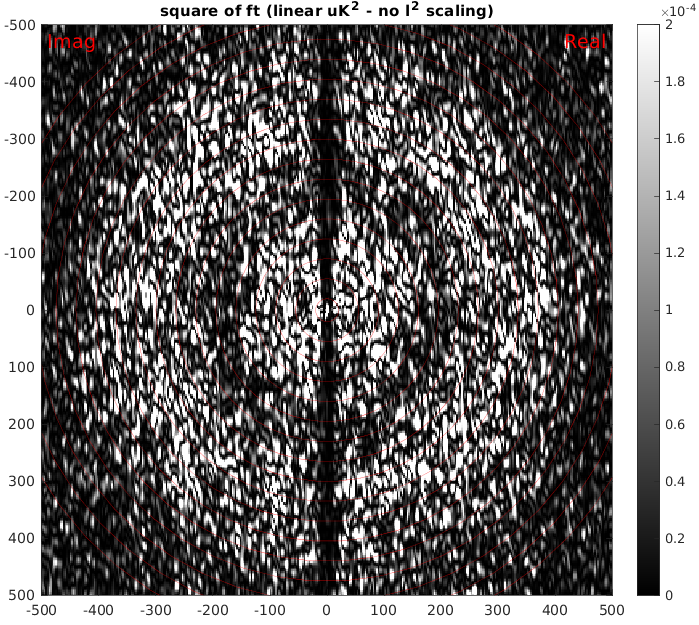}
			\includegraphics[width=0.4\textwidth]{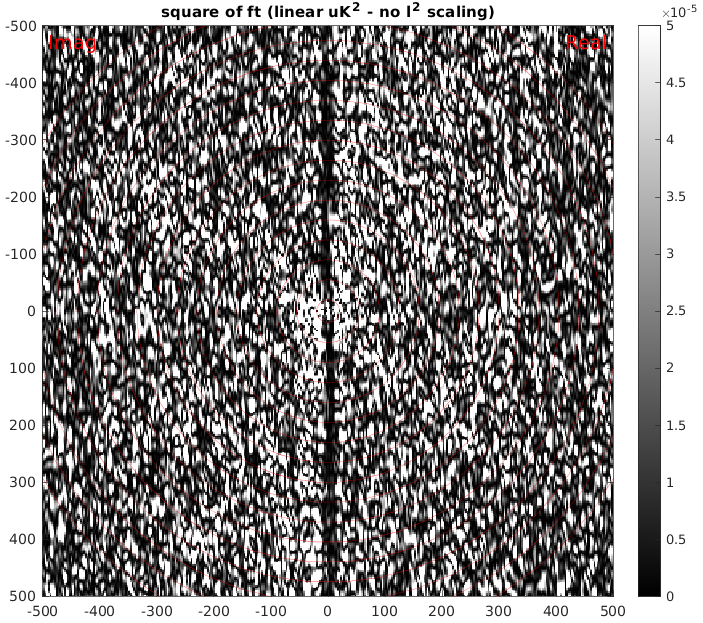}
		\end{tabular}
	\end{center}
	\caption[example] 
	{ \label{fig:B2019CS-II_eb_2daps} E-mode (\textit{Left}) and B-mode (\textit{Right}) 2D angular power spectra from late 2019 season Cold Spot observations\cite{Kang_thesis}. The depression along the $\ell_y$-axis is due to large scale filters to remove $1/f$ noise and ground-fixed signals. The two peaks in the E-mode spectrum are visible, while the B-mode spectrum is noise-dominated. The color axes are in different scales for E and B modes.}
	\end{figure}
	\begin{figure}[tb]
		\centering
		\includegraphics[width=0.45\textwidth]{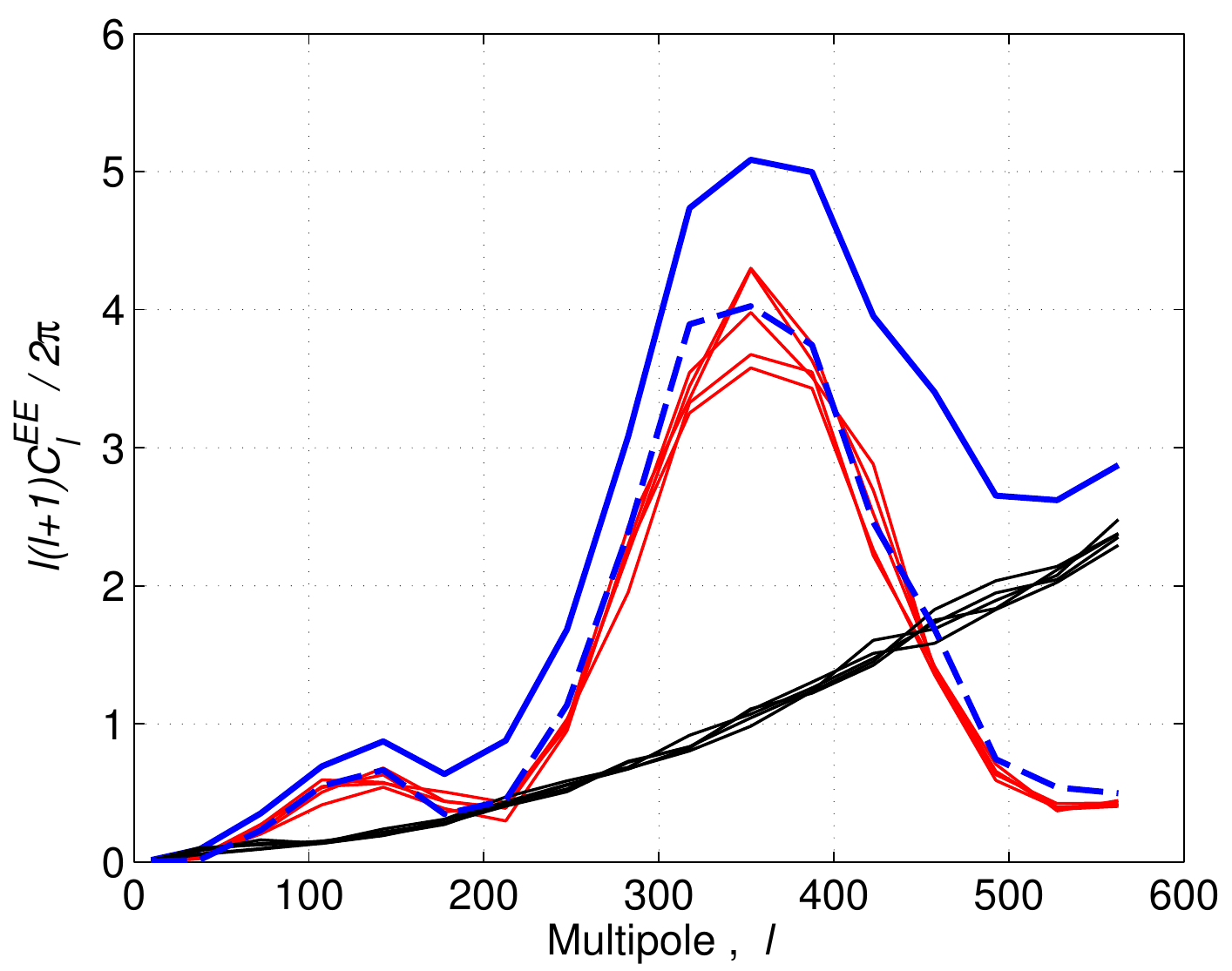}
		\includegraphics[width=0.45\textwidth]{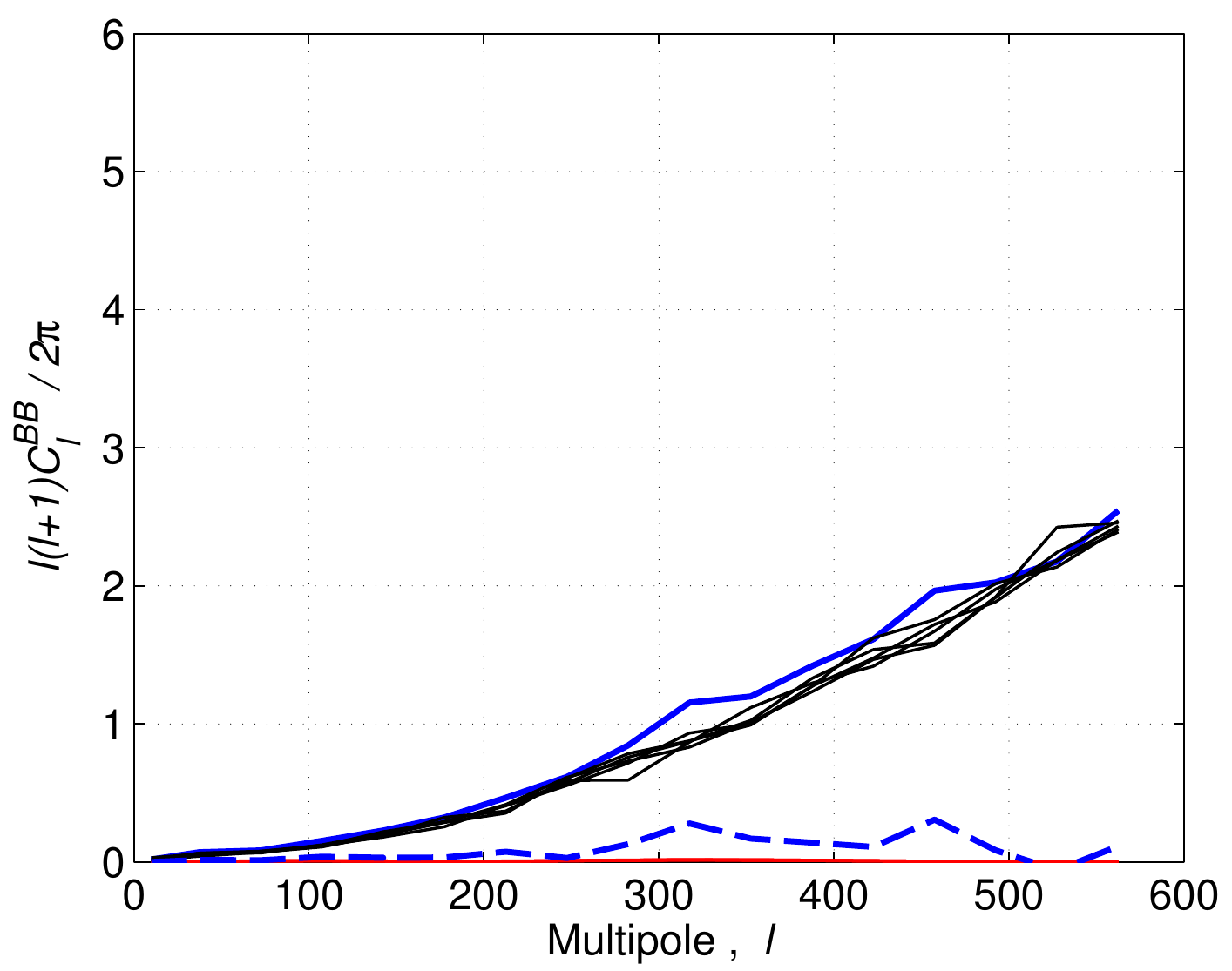}
		\caption[EE power spectrum from B3CS2019-II]{E-mode (\textit{Left}) and B-mode (\textit{Right}) 1D angular power spectra from late 2019 season Cold Spot observations\cite{Kang_thesis}. The solid blue curves are from the real observed data, the red solid curves are simulated observation of five unlensed $\Lambda$CDM signal realizations, and the black solid curves are from sign-flip noise realizations. The dashed blue curves are the real observed data minus the mean noise power. The suppression factors are not applied. We see clear detection of the E mode power spectrum at these angular scales consistent with $\Lambda$CDM cosmology.}
	\label{fig.aps_eb}
	\end{figure}

	Figures \ref{fig:B2019CS-II_ebmap}-\ref{fig.aps_eb} present the results from the late 2019 season Cold Spot observations\cite{Kang_thesis}. Figure~\ref{fig:B2019CS-II_ebmap} shows the E-mode and B-mode maps and their noise estimates, representing the observed modes within $50<\ell<120$. Degree scale E-modes are detected at high signal-to-noise, while the B-modes are noise-dominated. Figure \ref{fig:B2019CS-II_eb_2daps} shows the E-mode and B-mode 2D angular power spectra. The depression along the $\ell_y$-axis is due to large scale filters to remove $1/f$ noise and ground-fixed signals. The two peaks in the E-mode spectrum are visible, while the B-mdoe spectrum is noise-dominated. The color axes are in different scales for E and B modes. We bin the Fourier modes along the annuli to form 1D angular power spectra, shown in Figure \ref{fig.aps_eb}. The solid blue curves are the real observed data, the red solid curves are the simulated observation of five unlensed $\Lambda$CDM signal realizations, and the black solid curves are five noise realizations obtained by the sign-flip sequence technique\cite{BK-I}. The dashed blue curves are the real observed data minus the mean noise power. The suppression factors are not applied. We have shown five realizations of input signal and noise that are available at the moment, but we usually generate 499 realizations for full power spectrum analysis\cite{BK-X_BK15}. We already see a clear detection of the E mode power spectrum at these angular scales consistent with $\Lambda$CDM cosmology.
	
	\section{CONCLUSION}
	In this proceeding, we present the detection of E-modes at high signal-to-noise from BICEP3 low-elevation sky observations from the South Pole during 2019 summer seasons. The extended coverage was obtained with a flat mirror to redirect beams and was focused around the CMB Cold Spot to test the polarization anomaly. We detected standard $\Lambda$CDM E-mode statistics in the observed patch. The investigation around the Cold Spot may require a more sensitive polarization map to probe the polarization anomaly. With the detectors optimized for regular CMB observation, the map sensitivity above the declination of the Cold Spot is limited. The discriminating power from this observation may be weaker than expected by Ref.\citenum{ColdSpot_FC13} due to various factors: partial coverage around the Cold Spot, loss of large scale modes by timestream filters, and large noise around the edge of the observing field near the Cold Spot. Remaining analysis tasks for the map data around the Cold Spot include jackknife tests to probe potential systematics and proper weighting of noisy map pixels for the hypothesis test. A design of baffling that can work with the flat mirror is being studied for potential future observation. BICEP3 has enormous potential for a wide variety of E-mode science applications at degree angular scales from the South Pole.
	
	\acknowledgments 
	The BICEP/Keck project (including BICEP2, BICEP3 and BICEP Array) have been made possible through a series of grants from the National Science Foundation including 0742818, 0742592, 1044978, 1110087, 1145172, 1145143, 1145248, 1639040, 1638957, 1638978, 1638970, 1726917, 1313010, 1313062, 1313158, 1313287, 0960243, 1836010, 1056465, \& 1255358 and by the Keck Foundation. The development of antenna-coupled detector technology
	was supported by the JPL Research and Technology Development Fund and NASA Grants 06-ARPA206-0040, 10-SAT10-0017, 12-SAT12-0031, 14-SAT14-0009, 16-SAT16-0002, \& 18-SAT18-0017. The development and testing of focal planes were supported by the Gordon and Betty Moore Foundation at Caltech. Readout electronics were supported by a Canada Foundation for Innovation grant to UBC. The computations in this paper were run on the Odyssey cluster supported by the FAS Science Division Research Computing Group at
	Harvard University. The analysis effort at Stanford and SLAC was partially supported by the Department of Energy, Contract DE-AC02-76SF00515. We thank the staff of the U.S. Antarctic Program and in particular	the South Pole Station without whose help this research would not have been possible. Tireless administrative support was provided by Kathy Deniston, Sheri Stoll, Irene Coyle, Donna Hernandez, and Dana Volponi, and Julie Shih.
	
	\bibliography{report} 
	\bibliographystyle{spiebib} 
	
\end{document}